\documentclass[aps,prb,twocolumn,showpacs,preprintnumbers,prb,superscriptaddress]{revtex4-2}
\usepackage{graphicx}  
\usepackage{dcolumn}   
\usepackage{bm}        
\usepackage{amssymb}   
\usepackage{framed}
\usepackage{amsmath}
\usepackage{hhline}
\usepackage{xcolor}
\definecolor{mygreen}{RGB}{60,150,60}
\usepackage{amsmath,verbatim,moreverb}
\usepackage{tabularx}
\usepackage{lipsum}
\usepackage{threeparttable}
\usepackage{booktabs,amssymb,siunitx}
\usepackage{latexsym}
\usepackage{dcolumn}

\usepackage{subfig}
\usepackage{epsf}
\usepackage{float}
\usepackage{url}
\usepackage{hyperref}
\usepackage{caption}
\usepackage{xcolor}
\usepackage{tikz}
\usepackage[utf8]{inputenc}
\usepackage[normalem]{ulem}
\usepackage{ragged2e}
\usepackage{etoolbox}
\newcounter{subeqn} %

\usepackage{mathptmx}
\usepackage{tikz}
\usetikzlibrary{arrows.meta, calc, decorations.pathmorphing}
\usepackage{amsmath}
\usepackage{amssymb}
\usepackage[T1]{fontenc}
\usepackage{booktabs}
\usepackage{tabularx}
\usepackage{adjustbox}

\usepackage{hyperref}
\hypersetup{
    unicode=true,          
    pdftoolbar=true,        
    pdfmenubar=true,        
    pdffitwindow=false,     
    pdfstartview={FitH},    
    colorlinks=true,       
    linkcolor=blue,          
    citecolor=blue,        
    urlcolor=blue,
}

\begin{document}

\title{
Symmetry-Engineered Nonlinear Hall Response and Optical Response in Strained Monolayer Janus AsTeBr
}
\author{Dimple Rani}
\email{dimple.rani@niser.ac.in}
\affiliation{School of Physical Sciences, National Institute of Science Education and Research, An OCC of Homi Bhabha National Institute, Jatni 752050, India}

\author{Ritesh Ranjan Badhai}
\email{riteshranjanbadhai@gmail.com}
\affiliation{School of Physical Sciences, National Institute of Science Education and Research, An OCC of Homi Bhabha National Institute, Jatni 752050, India}

\author{Gayatri Panda}
\affiliation{Department of Chemistry and Pharmaceutical Sciences, Vrije Universiteit, Amsterdam, The Netherlands}

\author{Subrata Jana}
 \affiliation{Institute of Physics, Faculty of Physics, Astronomy and Informatics, Nicolaus Copernicus University in Toru\'n,
ul. Grudzi\k{a}dzka 5, 87-100 Toru\'n, Poland}

\author{Prasanjit Samal}
\affiliation{School of Physical Sciences, National Institute of Science Education and Research, An OCC of Homi Bhabha National Institute, Jatni 752050, India}

\begin{abstract}
The nonlinear Hall effect (NLHE) enables the generation of a transverse charge current in nonmagnetic materials with broken inversion symmetry while preserving time-reversal symmetry through the Berry curvature dipole (BCD). However, in crystals with $C_{3v}$ symmetry, the threefold rotational symmetry forces the BCD to vanish, thereby suppressing the intrinsic NLHE despite the presence of finite local Berry curvature. Here, using first-principles density functional theory combined with Wannier-based transport calculations, we demonstrate that uniaxial strain induces the NLHE in monolayer Janus AsTeBr by lowering the crystal symmetry from $C_{3v}$ to $C_{1}$ and generating a finite BCD. The resulting anisotropic redistribution of Berry-curvature hotspots produces pronounced nonlinear Hall conductivity and nonlinear Hall current, with the maximum response obtained at 2\% tensile strain. To elucidate the accompanying electronic-structure evolution, we further investigate the strain-dependent optical properties through the joint density of states, dielectric function, optical absorption, and reflectance. The optical spectra exhibit a systematic red shift and enhanced low-energy interband transitions, consistent with the strain-induced reconstruction of the electronic structure. Our results establish a microscopic connection between symmetry breaking, Berry-phase geometry, nonlinear Hall transport, and optical response, demonstrating that uniaxial strain provides an effective strategy for tailoring multiple functional properties in Janus two-dimensional materials.
\end{abstract}

\maketitle

\section{Introduction}
Since the isolation of graphene, two-dimensional (2D) materials have emerged as an exceptional platform for exploring novel quantum phenomena owing to their reduced dimensionality, tunable electronic structures, and strong light--matter interactions \cite{Novoselov2004,Geim2007,7plx-jqlw,8vvn-k9p3}. Beyond graphene, layered materials such as transition-metal dichalcogenides (TMDCs), black phosphorus, hexagonal boron nitride, MXenes, silicides, and magnetic van der Waals crystals exhibit diverse electronic, optical, magnetic, and mechanical properties, enabling applications in nanoelectronics, spintronics, valleytronics, photocatalysis, sensing, and optoelectronics \cite{Mak2010,Splendiani2010,Xu2014,Manzeli2017,Qiao2014,Naguib2011,Chhowalla2016,jj81-ng8g,doi:10.1021/acsanm.5c05785}. Among them, Janus monolayers form a unique class of intrinsically noncentrosymmetric materials, created by replacing one atomic layer of a conventional 2D crystal with a chemically distinct element, thereby breaking inversion symmetry ($\mathcal{I}$) while preserving the underlying lattice \cite{Lu2017,Dong2017}. Since the experimental realization of Janus MoSSe, this family has rapidly expanded to include TMDCs such as MoSSe, WSSe, MoSTe, and WSeTe, magnetic Janus trihalides Cr$_2$X$_3$Y$_3$ (X, Y = Cl, Br, I; X $\neq$ Y), Mn$_2$X$_3$Y$_3$ (X, Y = Cl, Br, I; X $\neq$ Y), CrXY (X, Y = S, Se, Te, Cl, Br, I; X $\neq$ Y), and Janus transition-metal chalcogenides such as CrXTe (X = S, Se), illustrating the rapidly growing chemical diversity of Janus materials beyond conventional TMDCs \cite{Lu2017,Zhang2017,Dong2017,jan,janus,crjan}. Their broken $\mathcal{I}$ symmetry, combined with the intrinsic out-of-plane electric field and strong spin--orbit coupling (SOC), gives rise to Rashba spin splitting, enhanced piezoelectricity, ferroelasticity, nonlinear optical responses, and Berry-phase-mediated transport, making Janus materials promising candidates for spintronic, valleytronic, nonlinear electronic, and optoelectronic applications \cite{Xiao2012,Xu2014,Mak2018,Zheng2020,PhysRevB.57.1505,9wgk-g3dd,doi:10.1021/acs.jpcc.4c07037}.

Monolayer Janus AsTeBr represents a particularly attractive member of this family because it combines intrinsic $\mathcal{I}$-symmetry breaking, strong SOC arising from the heavy Te and Br atoms, and excellent structural stability within a nonmagnetic semiconducting ground state \cite{10.1039/d3ta01177a,AsTeBr2}. Previous theoretical studies have established its crystal structure, electronic band dispersion, and optical characteristics, demonstrating its potential for electronic and optoelectronic applications \cite{AsTeBr2, Kocaba}. Nevertheless, the Berry-phase-mediated transport properties of AsTeBr have received little attention. In particular, although the absence of $\mathcal{I}$ symmetry allows finite local Berry curvature while preserving time-reversal symmetry ($\mathcal{T}$), it remains unclear whether the underlying crystal symmetry permits the emergence of experimentally measurable nonlinear transport and how such transport is coupled to the strain-dependent evolution of its electronic and optical properties.

A direct consequence of Berry curvature in noncentrosymmetric crystals is the nonlinear Hall effect (NLHE), which enables the generation of a transverse Hall current in materials preserving $\mathcal{T}$ symmetry without requiring either spontaneous magnetization or an external magnetic field \cite{Sodemann2015,Ma2019,Kang2019}. Unlike the conventional anomalous Hall effect, the NLHE originates from the Berry curvature dipole (BCD), corresponding to the first moment of the Berry curvature over the occupied states in momentum space \cite{Sodemann2015}. Owing to its quantum geometric origin, the NLHE has emerged as a powerful probe of Berry-phase physics and crystal symmetry, stimulating extensive theoretical and experimental studies in Weyl semimetals, topological materials, transition-metal dichalcogenides, and other inversion-asymmetric systems \cite{Xu2018,You2018,He2021}. However, the existence of a finite BCD is governed entirely by crystal symmetry. For crystals belonging to the $C_{3v}$ point group, the threefold rotational symmetry enforces an exact cancellation of the BCD, thereby suppressing the intrinsic nonlinear Hall response despite the presence of finite local Berry curvature \cite{Sodemann2015}. Since monolayer AsTeBr crystallizes in the $C_{3v}$ point group, its nonlinear Hall response is symmetry forbidden in the pristine state, motivating the search for an effective strategy to lift this symmetry constraint.

Among the available approaches, strain engineering has emerged as one of the most versatile and experimentally accessible methods for tailoring the electronic structure and symmetry of two-dimensional materials. In particular, uniaxial strain lowers the crystal symmetry while preserving $\mathcal{T}$ symmetry, thereby redistributing the Berry curvature in momentum space and enabling the emergence of a finite Berry curvature dipole and nonlinear Hall response \cite{Du2021,He2021,Liao2026}. Simultaneously, strain modifies orbital hybridization and interband transition energies, resulting in pronounced changes in the dielectric response, optical absorption, reflectance, and other optoelectronic properties \cite{Qian2014,Wang2018}. These concurrent modifications suggest that strain engineering provides a unified route for simultaneously controlling Berry-curvature-driven transport and optical functionalities in Janus materials.

In this work, we employ first-principles density functional theory combined with Wannier-based transport calculations to investigate the influence of uniaxial strain on the nonlinear Hall and optical responses of monolayer Janus AsTeBr. We demonstrate that symmetry lowering induced by uniaxial strain lifts the symmetry-enforced cancellation of the Berry curvature dipole, leading to the emergence of finite nonlinear Hall conductivity and nonlinear Hall current. Furthermore, we establish a direct correlation between the strain-induced evolution of the electronic structure and the accompanying changes in the joint density of states, dielectric response, optical absorption, and reflectance. Our results reveal symmetry engineering as an effective strategy for simultaneously manipulating nonlinear transport and optical functionalities in Janus materials, providing valuable design principles for future spintronic and optoelectronic devices.
\section{\label{meth}Computational Methodology}

First-principles calculations were carried out within the framework of density functional
theory (DFT) as implemented in the Vienna \textit{Ab initio} Simulation Package
(\textsc{vasp})~\cite{kresse1996efficiency}. The electron ion interactions are described using the projector augmented wave (PAW) method~\cite{kresse1996efficient,kresse1999ultrasoft}, 
and the exchange correlation energy is treated within the generalized gradient approximation (GGA) of Perdew, Burke, and Ernzerhof (PBE)~\cite{perdew1996generalized}.  A plane-wave energy cutoff of
520~eV was employed for expansion of the Kohn--Sham wave functions. The Brillouin zone
(BZ) was sampled using a $12 \times 12 \times 1$ $\Gamma$-centered Monkhorst-Pack
$k$-point mesh~\cite{Monkhorst1976}. 
Structural relaxations were performed until the total energy and Hellmann--Feynman forces
converged to within $10^{-6}$~eV and $0.001$~eV/\AA{}, respectively. SOC was included self-consistently throughout, given the presence of the heavy As and Br
atoms and its expected significant influence on the Berry curvature distribution of monolayer
AsTeBr.  Phonon dispersions are computed using the \textsc{PHONOPY} code~\cite{TOGO20151} within the finite-displacement approach 
using $4 \times 4$ supercells to verify the dynamical stability of all structures.

To enable efficient and accurate evaluation of Berry-phase-related quantities across the
full Brillouin zone, the converged DFT Bloch states of monolayer AsTeBr were used to
construct maximally localized Wannier functions (MLWFs)~\cite{marzari1997maximally,souza2001maximally} using the \textsc{wannier90}
package~\cite{pizzi2020wannier90}. The
Wannier-interpolated Hamiltonian, together with the associated position-operator matrix
elements, was subsequently used as input for the evaluation of the Berry curvature, the
Berry curvature dipole, and the nonlinear anomalous Hall conductivity using the
\textsc{WannierBerri} code~\cite{Tsirkin2021}. \textsc{WannierBerri} enables dense and
adaptively refined $k$-space integration, which is essential for capturing the sharply
peaked Berry-curvature contributions that arise near band degeneracies and saddle
points in the AsTeBr band structure.
For a Bloch band $n$ with periodic cell function $|u_n(\mathbf{k})\rangle$, the Berry
connection is defined as~\cite{RevModPhys.82.1959}
\begin{equation}
\mathbf{A}_n(\mathbf{k}) = i \langle u_n(\mathbf{k}) | \nabla_{\mathbf{k}} | u_n(\mathbf{k}) \rangle,
\label{eq:berry_connection}
\end{equation}
and the associated Berry curvature is
\begin{equation}
\boldsymbol{\Omega}_n(\mathbf{k}) = \nabla_{\mathbf{k}} \times \mathbf{A}_n(\mathbf{k}).
\label{eq:berry_curvature}
\end{equation}
The Berry curvature acts as an effective magnetic field in momentum space and gives rise to an anomalous transverse velocity of Bloch electrons under an external electric field. Within the Kubo linear-response formalism, the Berry curvature of the $n$th band is evaluated as~\cite{yao2004first,RevModPhys.82.1959}

\begin{equation}
\Omega^{z}_{n}(\mathbf{k})
=
-2\hbar^{2}
\sum_{m\neq n}
\frac{
\mathrm{Im}
\left[
\langle n|\hat{v}_{x}|m\rangle
\langle m|\hat{v}_{y}|n\rangle
\right]
}
{\left(E_{m}-E_{n}\right)^{2}},
\label{eq:kubo_bc}
\end{equation}
where $\hat v_x$ and $\hat v_y$ denote the velocity operators along the $x$ and $y$ directions, respectively, and $E_n$ is the energy of the $n$th Bloch band. Since monolayer AsTeBr is a two-dimensional system, only the out-of-plane component $\Omega_n^{z}(\mathbf{k})$ is finite and contributes to the transport properties.
For systems preserving time-reversal symmetry ($\mathcal{T}$), the Berry curvature satisfies\cite{RevModPhys.82.1959}

\begin{equation}
\Omega_n(\mathbf{k})
=
-
\Omega_n(-\mathbf{k}),
\label{eq:TRS}
\end{equation}
whereas inversion symmetry ($\mathcal{I}$) imposes
\begin{equation}
\Omega_n(\mathbf{k})
=
\Omega_n(-\mathbf{k}).
\label{eq:IS}
\end{equation}

Consequently, when both $\mathcal{T}$ and $\mathcal{I}$ are present, the Berry curvature vanishes identically throughout the Brillouin zone. In Janus AsTeBr, the inequivalent Te and Br atomic layers break inversion symmetry while preserving time-reversal symmetry, allowing finite local Berry curvature to develop around the valleys. Nevertheless, the intrinsic linear anomalous Hall conductivity remains zero because the Berry curvature is an odd function of momentum and its Brillouin-zone integral cancels exactly.

The leading Hall response therefore appears at second order in the applied electric field and is characterized by the Berry curvature dipole (BCD), defined as~\cite{Sodemann2015}

\begin{equation}
D_{ab}
=
\sum_n
\int
\frac{d^2k}{(2\pi)^2}
f_n(\mathbf{k})
\frac{\partial \Omega_n^{b}(\mathbf{k})}{\partial k_a},
\label{eq:bcd}
\end{equation}
where $f_n(\mathbf{k})$ is the equilibrium Fermi--Dirac distribution function, while the indices $a$ and $b$ denote the momentum derivative and Berry-curvature directions, respectively. Integrating Eq.\eqref{eq:bcd} by parts yields

\begin{equation}
D_{ab}
=
-
\sum_n
\int
\frac{d^2k}{(2\pi)^2}
\frac{\partial f_n(\mathbf{k})}{\partial k_a}
\Omega_n^{b}(\mathbf{k}),
\label{eq:bcd_parts}
\end{equation}
which explicitly shows that the Berry curvature dipole originates from partially occupied electronic states in the vicinity of the Fermi level.

Although finite local Berry curvature exists in pristine AsTeBr, its crystal belongs to the noncentrosymmetric $C_{3v}$ point group. The threefold rotational symmetry enforces an exact cancellation of the first moment of the Berry curvature over the Brillouin zone, resulting in a vanishing BCD. Application of uniaxial strain lowers the crystal symmetry, removes this symmetry constraint, and consequently generates a finite Berry curvature dipole, giving rise to the nonlinear Hall response discussed below. Under an oscillating electric field,
$\mathbf{E}(t)=\mathrm{Re}\left[\mathbf{E}(\omega)e^{-i\omega t}\right],$
the semiclassical dynamics of Bloch electrons can be described within the Boltzmann transport formalism under the relaxation-time approximation~\cite{ziman2001electrons}. The second-order Hall current induced by the Berry-curvature dipole is expressed as

\begin{equation}
j_a^{(2\omega)}
=
\chi_{abc}
E_b(\omega)
E_c(\omega),
\label{eq:NLHC}
\end{equation}
where $\chi_{abc}$ is the nonlinear Hall conductivity tensor, and the indices $a$, $b$, and $c$ denote the current and electric-field directions. Following the semiclassical theory developed by Sodemann and Fu~\cite{Sodemann2015}, the nonlinear Hall conductivity is directly related to the Berry curvature dipole through

\begin{equation}
\chi_{abc}
=
\varepsilon_{acd}
\frac{e^{3}\tau}
{2\left(1+i\omega\tau\right)}
D_{bd},
\label{eq:chi}
\end{equation}
where $\varepsilon_{acd}$ is the Levi--Civita antisymmetric tensor, $e$ is the elementary charge, $\tau$ denotes the carrier relaxation time, and $\omega$ is the frequency of the applied electric field. The Berry curvature dipole tensor $D_{bd}$ couples the electric field to the nonlinear Hall current, while the Levi--Civita tensor ensures that the induced current is transverse to the applied field.

In the low-frequency limit ($\omega\tau\ll1$), corresponding to the dc transport regime, Eq.~\ref{eq:chi} reduces to

\begin{equation}
\chi_{abc}
=
\varepsilon_{acd}
\frac{e^{3}\tau}{2}
D_{bd},
\label{eq:chi_dc}
\end{equation}
demonstrating that the nonlinear Hall conductivity is directly proportional to the Berry curvature dipole. Consequently, the magnitude and symmetry of the nonlinear Hall response are completely determined by the symmetry-allowed components of $D_{ab}$, whereas the relaxation time acts only as an overall scaling factor. 

The activation of the nonlinear Hall response under uniaxial strain originates from the strain-induced modification of the electronic band structure and the associated redistribution of the Berry curvature near the Fermi level. These changes not only determine the Berry curvature dipole responsible for the nonlinear Hall effect but also alter the interband optical transitions. Consequently, the optical response provides complementary information on the microscopic electronic structure underlying the nonlinear transport behavior. To establish this correlation, we further computed four frequency-dependent optical quantities: the joint density of states (JDOS), the imaginary part of the dielectric function $\mathrm{Im}[\varepsilon(\omega)]$, the absorption coefficient, and the reflectivity spectrum.

Within the independent-particle approximation, the frequency-dependent dielectric response is described by the complex dielectric function~\cite{gajdos2006linear,Wiser1963}

\begin{equation}
\varepsilon(\omega)=\varepsilon_{1}(\omega)+i\varepsilon_{2}(\omega),
\end{equation}
where $\varepsilon_{1}(\omega)$ and $\varepsilon_{2}(\omega)$ denote the real and imaginary parts of the dielectric function, respectively. Since the imaginary component originates directly from optical transitions between occupied and unoccupied electronic states, it serves as the fundamental quantity from which the remaining optical constants are derived.  the dielectric function is evaluated as~\cite{gajdos2006linear}

\begin{equation}
\begin{aligned}
\varepsilon_{2}^{\alpha\beta}(\omega)
&=
\frac{4\pi^{2}e^{2}}{\Omega}
\lim_{q\rightarrow0}
\frac{1}{q^{2}}
\sum_{c,v,\mathbf{k}}
2w_{\mathbf{k}}
\delta
(E_{c\mathbf{k}}-E_{v\mathbf{k}}-\hbar\omega)
\\
&\qquad\times
\left|
\left<
u_{c,\mathbf{k}+\mathbf{e}_{\alpha}q}
\middle|
u_{v,\mathbf{k}}
\right>
\right|^{2}.
\end{aligned}
\end{equation}
where $\Omega$ is the unit-cell volume, $w_{\mathbf{k}}$ is the Brillouin-zone weighting factor, $v$ and $c$ label the occupied valence and unoccupied conduction bands, respectively, and $\mathbf{e}_{\alpha}$ denotes the polarization direction of the incident electric field. The real part of the dielectric function was subsequently obtained through the Kramers--Kronig transformation~\cite{yu2010fundamentals}.

To identify the microscopic origin of the optical spectra, the joint density of states was also evaluated as\cite{yu2010fundamentals}

\begin{equation}
J_{\mathrm{JDOS}}(\omega)
=
\sum_{v,c}
\int_{\mathrm{BZ}}
\frac{d^{2}k}{(2\pi)^{2}}
\delta
(E_{c\mathbf{k}}-E_{v\mathbf{k}}-\hbar\omega),
\end{equation}
which measures the number of energetically allowed interband transitions at a given photon energy. Comparison between the JDOS and $\varepsilon_{2}(\omega)$ enables direct identification of the electronic transitions responsible for the prominent optical features.

The absorption coefficient and reflectivity were subsequently obtained from the complex dielectric function according to\cite{yu2010fundamentals}

\begin{equation}
\alpha(\omega)
=
\frac{\sqrt{2}\omega}{c}
\left[
\sqrt{\varepsilon_{1}^{2}(\omega)+\varepsilon_{2}^{2}(\omega)}
-
\varepsilon_{1}(\omega)
\right]^{1/2},
\end{equation}
and

\begin{equation}
R(\omega)
=
\left|
\frac{\sqrt{\varepsilon(\omega)}-1}
{\sqrt{\varepsilon(\omega)}+1}
\right|^{2},
\end{equation}
respectively. Together, these quantities provide a comprehensive description of the strain-dependent optical response and establish a direct connection between the evolution of the electronic structure, the interband optical transitions, and the nonlinear Hall transport in Janus AsTeBr.

\section{Results and Discussions}
\subsection{Crystal Symmetry and Electronic Structure}
\begin{figure}[ht]
    \centering
    \includegraphics[width=1\linewidth]{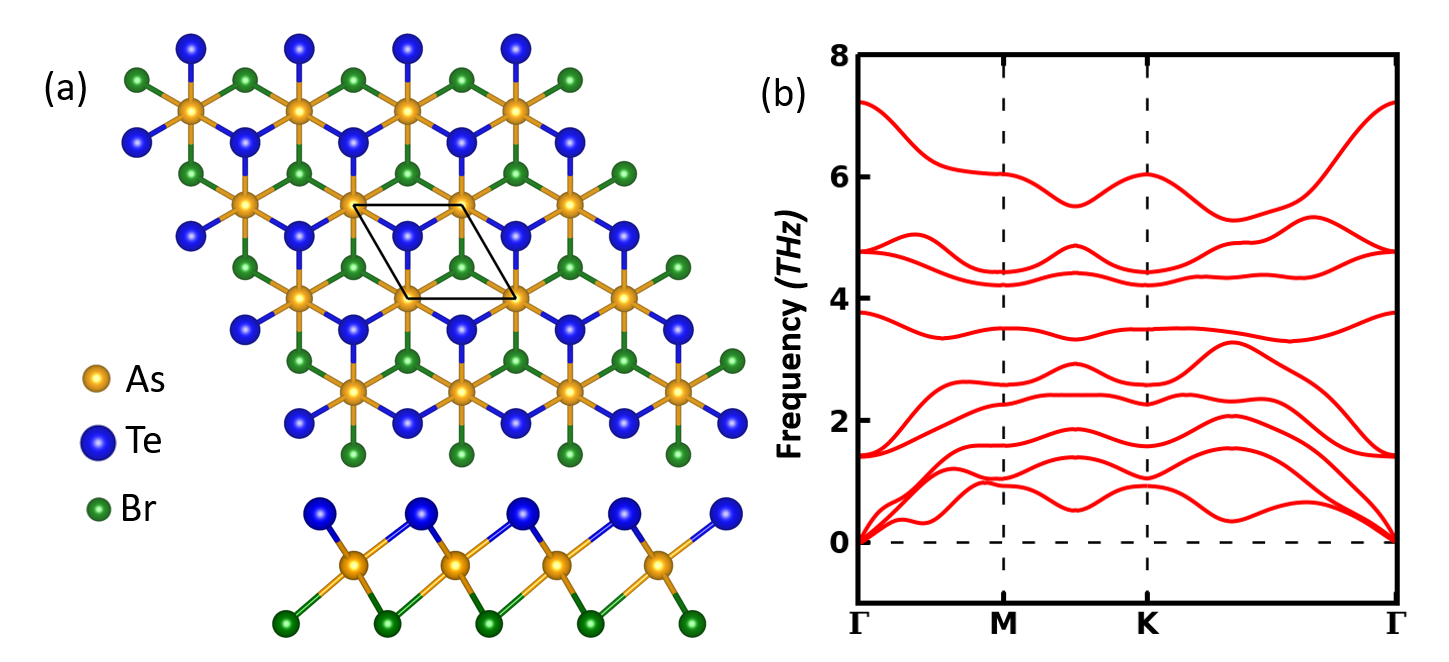}
    \caption{\justifying(a) Top and side views of the optimized crystal structure of the Janus monolayer AsTeBr. Gold, blue, and green spheres represent As, Te, and Br atoms, respectively. (b) Phonon dispersion along the high-symmetry path in the Brillouin zone. The absence of imaginary frequencies confirms the dynamical stability of the monolayer.}
    \label{fig1}
\end{figure}

\begin{figure*}
    \centering
    \includegraphics[width=1\linewidth]{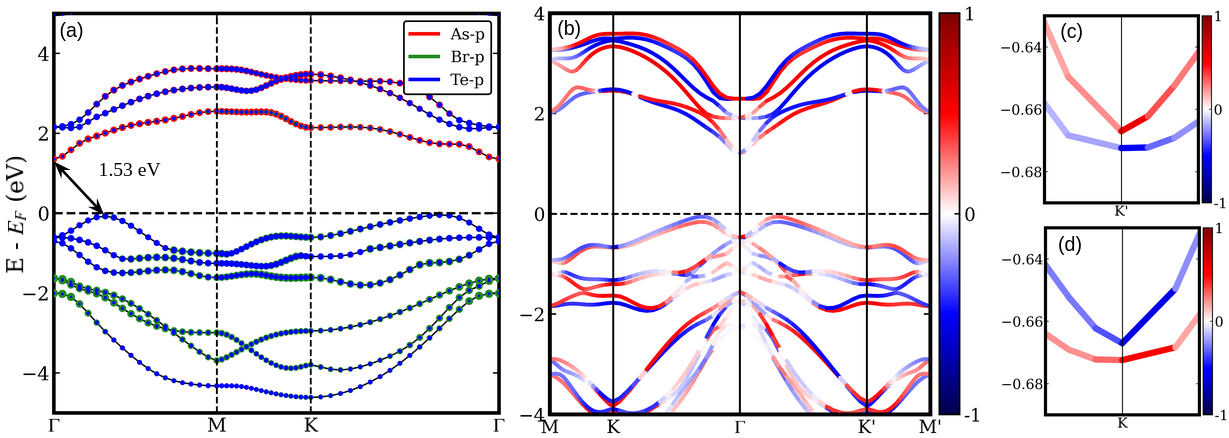}
   \caption{\justifying
(a) Orbital-projected electronic band structure of monolayer AsTeBr. The dashed line denotes the Fermi level ($E_F=0$), revealing an indirect band gap of 1.53 eV. (b) Spin-resolved band structure including SOC, with the color scale representing the out-of-plane spin polarization, $\langle S_z \rangle$. (c)-(d) Enlarged views near the K and K$^\prime$ valleys showing SOC-induced spin splitting and valley-contrasting spin polarization.
}
\label{bs11}
\end{figure*}

The optimized crystal structure of the Janus AsTeBr monolayer is shown in Fig.~\ref{fig1}(a). The system crystallizes in a trigonal lattice with $C_{3v}$ symmetry and a hexagonal unit cell containing three atoms per primitive cell~\cite{AsTeBr2}. In this configuration, the As atom is located between Te and Br layers, forming a vertically asymmetric arrangement that breaks the out-of-plane $\mathcal{I}$. The top and side views clearly illustrate the buckled geometry of the monolayer, where the difference in atomic species on either side leads to a non-centrosymmetric structure. The optimized lattice constants are found to be $a = b = 3.86~\text{\AA}$, with a vacuum spacing of $18.3~\text{\AA}$ introduced along the out-of-plane direction to avoid spurious interactions between periodic images. The calculated bond lengths are $2.80~\text{\AA}$ for As--Te and $2.95~\text{\AA}$ for As-Br, indicating a slight elongation of the As-Br bond compared to As-Te. This variation in bond lengths originates from the difference in atomic radii and electronegativity between Te and Br atoms, further contributing to the structural asymmetry of the Janus configuration.
The dynamical stability is verified by calculating the phonon dispersion, which shows no imaginary frequencies across the entire Brillouin zone as shown in Fig.~\ref{fig1} (b). These results demonstrate that the Janus AsTeBr monolayer is dynamically stable, indicating its feasibility for further investigation. 
Fig.~\ref{bs11}(a) shows the orbital-projected electronic band structure of monolayer Janus AsTeBr without SOC. The material is an indirect semiconductor with a band gap of approximately 1.53 eV. The valence-band maximum (VBM) is predominantly derived from Br-$p$ orbitals, whereas the conduction-band minimum (CBM) mainly consists of hybridized As-$p$ and Te-$p$ states. The distinct orbital contributions near the band edges originate from the inequivalent Te and Br atomic layers, reflecting the intrinsic structural asymmetry of the Janus lattice. Such orbital hybridization is expected to play an important role in the SOC-induced modification of the low-energy electronic structure. The SOC band structure is presented in Fig.~\ref{bs11}(b), where the color scale denotes the expectation value of the out-of-plane spin component, $\langle S_z\rangle$.

The inclusion of SOC lifts the spin degeneracy around the K and K$^\prime$ valleys owing to the broken $\mathcal{I}$ of the Janus structure. At the same time, the opposite spin polarization observed at the two inequivalent valleys reflects the preservation of $\mathcal{T}$. This valley-contrasting spin texture is a characteristic feature of noncentrosymmetric materials with strong SOC and indicates a strong coupling between the spin, orbital, and valley degrees of freedom.
The enlarged views around the K and K$^\prime$ valleys shown in Figs.~\ref{bs11}(c) and \ref{bs11}(d) clearly resolve the SOC-induced spin splitting near the band edges. The reversal of spin polarization between the two valleys is consistent with the time-reversal symmetry of the nonmagnetic ground state, whereas the finite spin splitting of 5.4 meV originates from the absence of inversion symmetry. The sizable valley-dependent spin splitting suggests that SOC plays a significant role in shaping the electronic structure near the Fermi level and is expected to strongly influence the Berry-curvature distribution and the resulting nonlinear transport properties discussed in the following sections.

\subsection{\label{bcdd}Strain-Modulated Berry Curvature and Berry Curvature Dipole}

The nonlinear Hall effect originates from the geometric properties of Bloch electrons encoded in the Berry curvature, which acts as an effective magnetic field in momentum space and governs the anomalous carrier velocity under an external electric field~\cite{Karplus1954,Xiao2010,Sodemann2015}. As discussed in Sec.~\ref{meth}, the Berry curvature is evaluated using the Kubo formalism [Eq.~(\ref{eq:kubo_bc})], while its symmetry is governed by Eqs.~(\ref{eq:TRS}) and (\ref{eq:IS}). Consequently, finite Berry curvature exists only in systems with broken  $\mathcal{I}$ symmetry\cite{Xiao2010}. Janus AsTeBr naturally satisfies this condition owing to the inequivalent Te and Br layers while preserving $\mathcal{T}$ symmetry. Nevertheless, the pristine monolayer belongs to the noncentrosymmetric $C_{3v}$ point group, whose threefold rotational symmetry enforces an exact cancellation of the first moment of the Berry-curvature distribution\cite{Sodemann2015,You2018}. As a result, although finite local Berry curvature is present, the BCD vanishes, precluding an intrinsic nonlinear Hall response.

To lift this symmetry constraint, we apply uniaxial strain. In contrast to biaxial strain, which preserves the $C_{3v}$ symmetry and therefore cannot generate a finite BCD, uniaxial strain lowers the crystal symmetry and enables a finite BCD. Uniaxial strain was examined independently along both the $x$ and $y$ directions, yielding nearly identical structural parameters and electronic band gaps; the corresponding results for strain applied along the $x$ direction are summarized in Table~S1 of the Supplemental Material~\cite{SM}. Furthermore, phonon calculations reveal that representative tensile-strained structures remain dynamically stable, whereas compressive strain exhibits imaginary phonon modes, indicating lattice instability (Fig.~S1, Supplemental Material ~\cite{SM}). Consequently, the following discussion focuses on tensile strain applied along the $x$ direction. This deformation lowers the point-group symmetry from $C_{3v}$ to $C_{1}$ while preserving the nonmagnetic ground state and $\mathcal{T}$ symmetry, thereby modifying the interband coupling responsible for the Berry curvature.
\begin{figure*}
    \centering
    \includegraphics[width=1\linewidth]{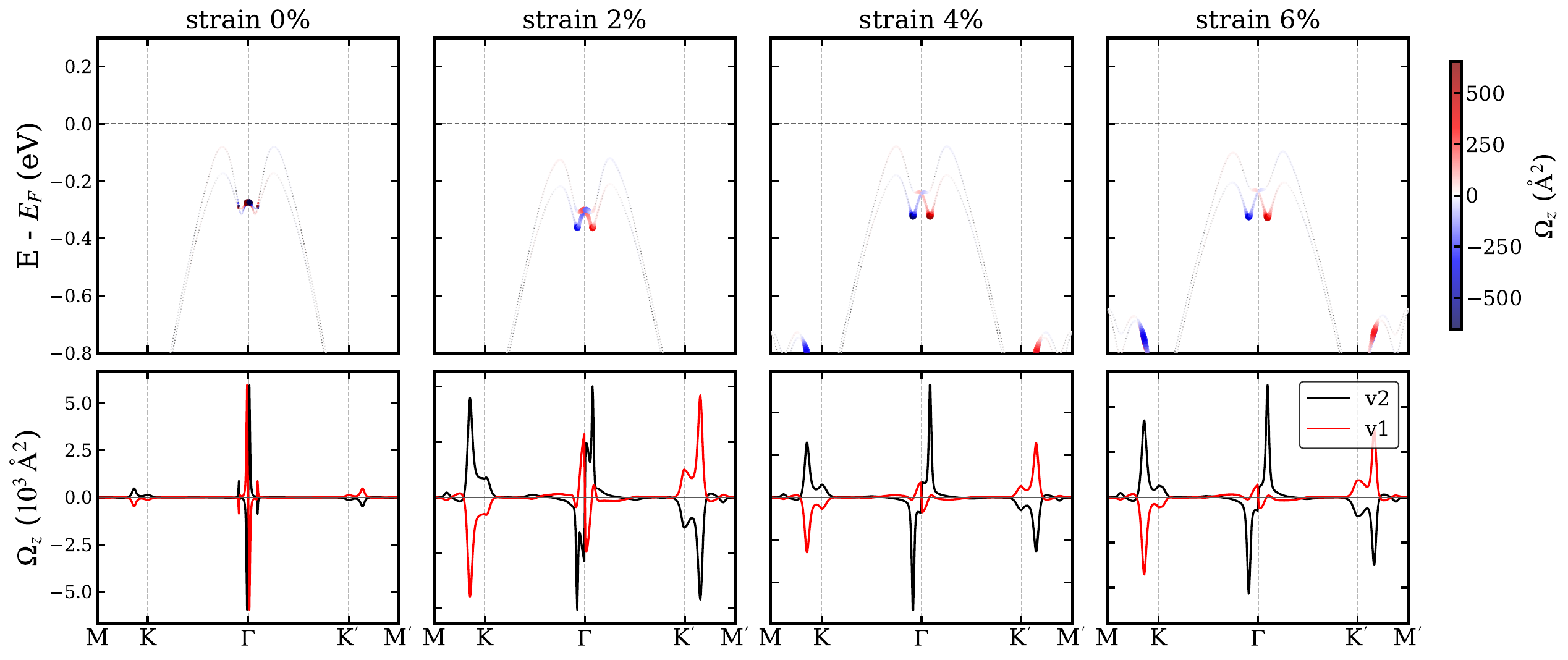}
    \caption{\justifying Band-resolved Berry curvature of monolayer Janus AsTeBr under 0\%, 2\%, 4\%, and 6\% uniaxial tensile strain. The top panels show the highest ($v_1$) and second-highest ($v_2$) valence bands with the Berry curvature projected onto the band structure, where the color and marker size denote the sign and magnitude of the Berry curvature, respectively. The bottom panels show the corresponding Berry curvature, $\Omega_z(k)$, of the $v_1$ (red) and $v_2$ (black) bands. }
    \label{fig:berry_band}
\end{figure*}
The strain-dependent evolution of the Berry curvature is presented in Fig.~\ref{fig:berry_band} shows the band-resolved Berry curvature of the two uppermost valence bands ($v_{1}$ and $v_{2}$). In the pristine monolayer, pronounced Berry-curvature hotspots are localized around the K and K$'$ valleys, where SOC lifts the near degeneracy of the valence bands and enhances interband coupling near avoided crossings. 

According to Eq.~(\ref{eq:kubo_bc}), the combination of small band separation and large interband velocity matrix elements produces strong Berry-curvature contributions with opposite signs at K and K$'$, consistent with the preserved $\mathcal{T}$ symmetry\cite{Xiao2007,Xiao2012}. Upon applying uniaxial strain, the modified crystal field and the evolution of the As-$p$, Te-$p$, and Br-$p$ orbital hybridization alter the band dispersion and the SOC-induced avoided crossings, leading to a continuous evolution of both the magnitude and spatial distribution of the Berry-curvature hotspots. Rather than creating Berry curvature, strain redistributes its momentum-space profile through changes in the electronic wave functions. The occupied Berry curvature, obtained by summing the contributions from all valence bands exhibits a similar strain-dependent evolution as shown in Fig. S2 of Supplementery material~\cite{SM}. 
\begin{figure} 
    \centering
    \includegraphics[width=1.0\linewidth]{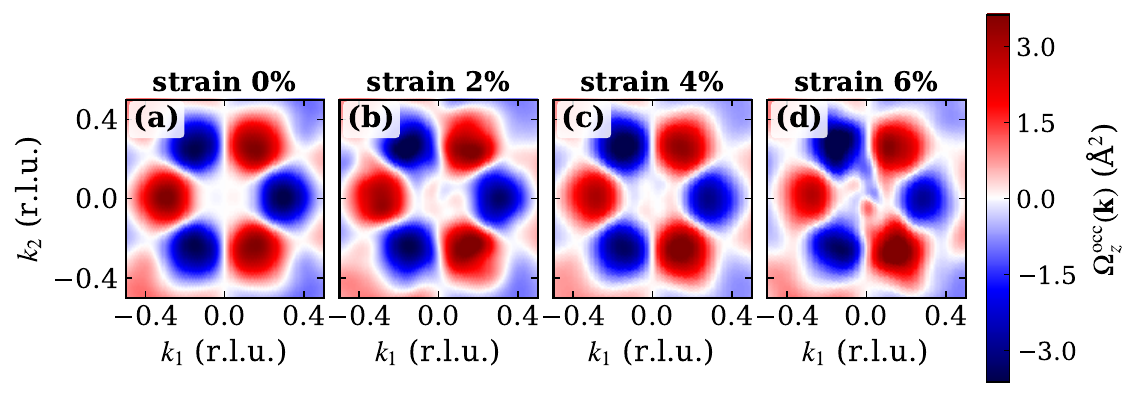}
    \caption{\justifying Occupied Berry curvature, $\Omega_z^{\rm occ}(\mathbf{k})$, of monolayer Janus AsTeBr in the first Brillouin zone under uniaxial tensile strain of (a) 0\%, (b) 2\%, (c) 4\%, and (d) 6\%.}
    \label{fig:omega2D}
\end{figure}
This evolution is more clearly visualized in the 2D Berry-curvature maps shown in Fig.~\ref{fig:omega2D}. The pristine monolayer exhibits the expected threefold rotational symmetry associated with the $C_{3v}$ crystal structure, whereas uniaxial strain progressively distorts the positive and negative Berry-curvature lobes without violating the antisymmetry imposed by $\mathcal{T}$. The resulting loss of rotational symmetry produces a directional asymmetry in momentum space, establishing the essential condition for the emergence of a finite BCD. 
\begin{figure} 
    \centering
    \includegraphics[width=1.0\linewidth]{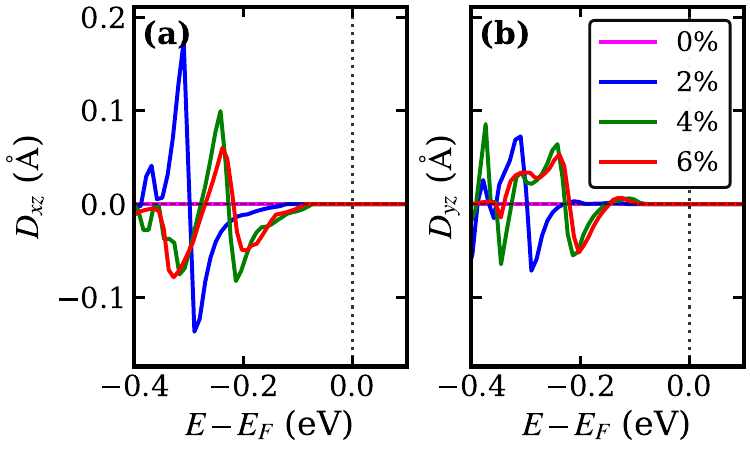}
   \caption{\justifying Energy-dependent BCD components $D_{xz}$ and $D_{yz}$ of monolayer Janus AsTeBr under 0--6\% uniaxial tensile strain. Panels (a) and (b) show the variation of $D_{xz}$ and $D_{yz}$, respectively, as a function of energy relative to the Fermi level.}
    \label{fig:bcd}
\end{figure} 

The nonlinear Hall effect is governed not by the Berry curvature itself but by its first moment, namely the Berry curvature dipole (BCD) defined in Eq.~(\ref{eq:bcd}). Fig.~\ref{fig:bcd} shows the energy dependence of the BCD components $D_{xz}$ and $D_{yz}$ under different uniaxial tensile strains (values given in Table S2 of supplementery material~\cite{SM}).  As expected from symmetry, both components remain negligible in the pristine monolayer over the entire energy range. Once the symmetry is reduced to $C_{1}$, finite BCD components emerge immediately, exhibiting pronounced energy dependence with the largest response occurring approximately 0.2~eV below the Fermi level. Among the investigated strains, the $D_{xz}$ component reaches its maximum magnitude at 2\% strain and remains considerably larger than $D_{yz}$, reflecting the anisotropic redistribution of the Berry curvature induced by uniaxial deformation. The sign reversals observed for both tensor components as the Fermi level is varied indicate opposite contributions from electron- and hole-like states and demonstrate the strong sensitivity of the BCD to the strain-modified electronic structure. The maximum Berry curvature dipole of $0.20477$~\AA\ exceeds the values reported for several representative 2D nonlinear Hall materials, including strained TMDC's and related low-symmetry compounds~\cite{xiong2025,BANDYOPADHYAY2024100101,maa2019,kangg2019}. This comparison highlights that uniaxial strain provides an efficient route for activating a sizable Berry curvature dipole in Janus AsTeBr through symmetry reduction from $C_{3v}$ to $C_{1}$.
\begin{figure}
    \centering
    \includegraphics[width=1.0\linewidth]{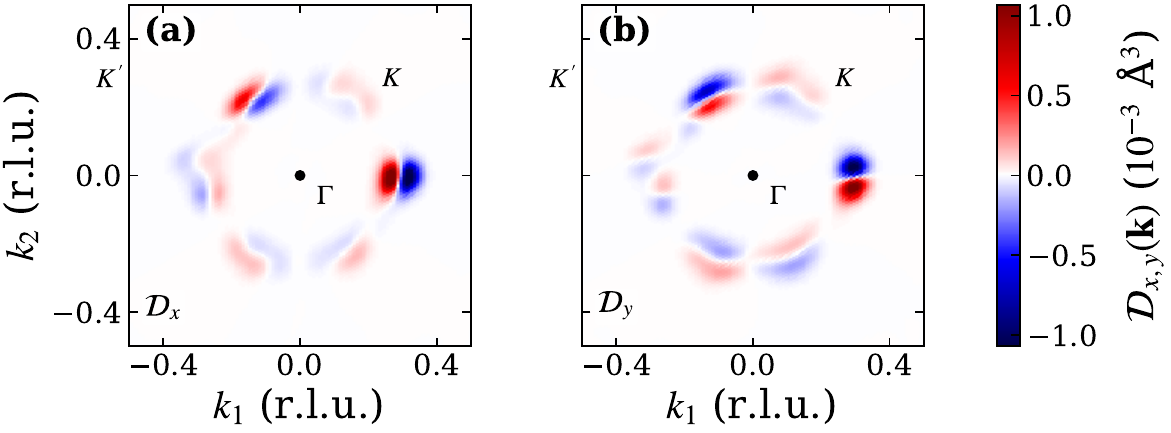}
    \caption{\justifying Momentum-space distribution of the occupied Berry curvature dipole of monolayer Janus AsTeBr under 2\% uniaxial tensile strain. (a)-(b) $k$-resolved Berry curvature dipole components, ${D}_x(\mathbf{k})$ and ${D}_y(\mathbf{k})$, evaluated at the Fermi level ($E_F=-2.28$ eV). The white dashed curves denote the Fermi surface.}
    \label{fig:bcd_map}
\end{figure}

The microscopic origin of the finite BCD is illustrated by the momentum-space distributions at 2\% tensile strain in Fig.~\ref{fig:bcd_map}. Corresponding distributions for 4\% and 6\% tensile strain are provided in Fig. S3 and S4 of the Supplemental Material~\cite{SM}. While the occupied Berry curvature remains distributed throughout the Brillouin zone as shwon in Fig.~\ref{fig:omega2D} (b), the corresponding $D_x(\mathbf{k})$ and $D_y(\mathbf{k})$ distributions from Fig.~\ref{fig:bcd_map}, reveal that the dominant contributions arise from regions surrounding the K and K$'$ valleys.  The strain-induced breaking of the threefold rotational symmetry produces an unequal weighting of positive and negative Berry-curvature contributions, preventing their complete cancellation and giving rise to finite dipole moments. According to the semiclassical theory of Sodemann and Fu\cite{Sodemann2015}, this strain-induced BCD directly generates the nonlinear Hall conductivity and nonlinear Hall current through Eq.~(\ref{eq:chi_dc}). 
\subsection{Nonlinear Hall Conductivity and Hall Current}
\begin{figure} [ht]
    \centering
    \includegraphics[width=1.0\linewidth]{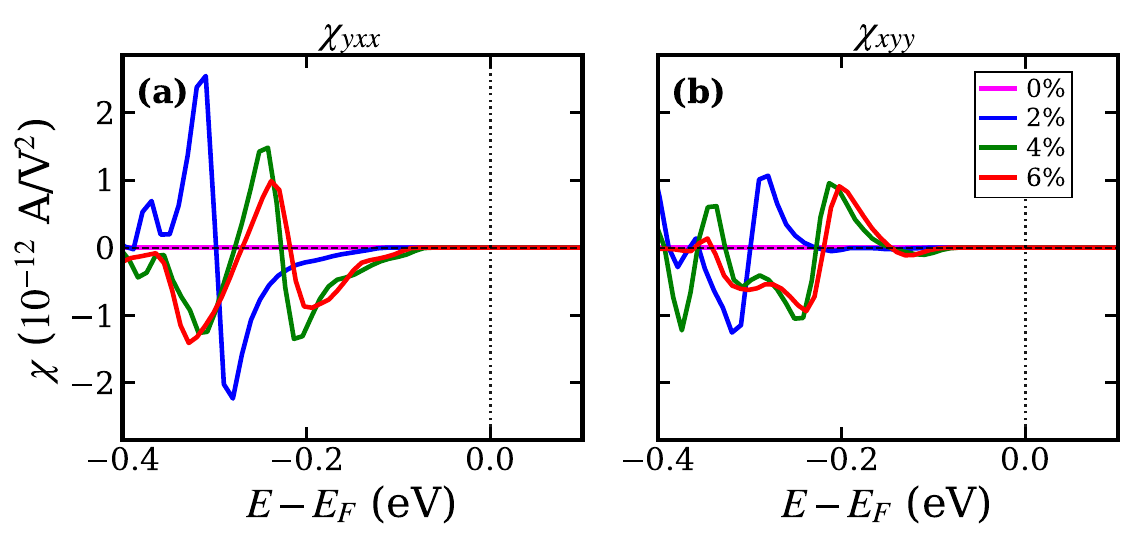}
\caption{\justifying Second-order nonlinear Hall conductivity tensor components (a) $\chi_{yxx}$ and (b) $\chi_{xyy}$ of monolayer Janus AsTeBr under uniaxial tensile strain. The vertical dotted line marks the intrinsic Fermi level.}
    \label{fig:nlhc}
\end{figure}
The strain-induced BCD discussed in ~\ref{bcdd} directly gives rise to a nonlinear Hall response through Eq.~(\ref{eq:chi_dc}). As shown in Fig.~\ref{fig:nlhc}, the calculated nonlinear Hall conductivity tensor components, $\chi_{yxx}$ and $\chi_{xyy}$, as a function of the Fermi energy under different uniaxial tensile strains. As expected from the vanishing Berry curvature dipole in the pristine monolayer, both tensor components remain essentially zero over the entire energy range. Once the crystal symmetry is lowered from $C_{3v}$ to $C_{1}$, finite nonlinear Hall conductivity emerges immediately, exhibiting pronounced energy dependence and multiple sign reversals. These features originate from the rapid variation of the BCD as the Fermi level traverses regions with different Berry-curvature contributions, leading to alternating electron- and hole-like contributions to the nonlinear Hall response.

The two conductivity tensor components exhibit distinct anisotropic behavior. The $\chi_{yxx}$ component displays substantially larger amplitudes than $\chi_{xyy}$, reaching its maximum value at approximately 2\% tensile strain, whereas larger strains produce broader but comparatively weaker peaks. This behavior indicates that the nonlinear Hall response is governed by a competition between symmetry breaking and strain-induced band reconstruction. Moderate strain efficiently removes the symmetry-enforced cancellation of the Berry curvature dipole while preserving strong SOC induced Berry-curvature hotspots near the Fermi level. At higher strains, further modification of the band dispersion redistributes these hotspots over a wider energy range, reducing the peak nonlinear Hall conductivity despite the stronger structural distortion. In contrast, the comparatively smaller magnitude of $\chi_{xyy}$ reflects the anisotropic redistribution of the Berry curvature induced by uniaxial deformation.
\begin{figure}
    \centering
    \includegraphics[width=1.0\linewidth]{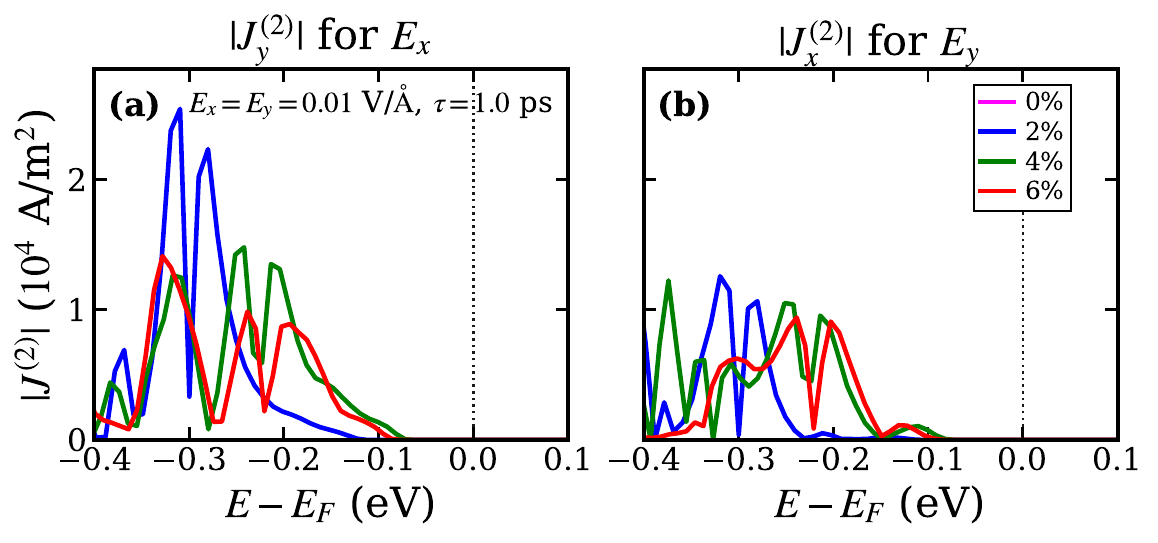}
    \caption{\justifying Magnitude of the second-order nonlinear Hall current density of monolayer Janus AsTeBr under uniaxial tensile strain. Panels (a) and (b) show $|J_y^{(2)}|$ induced by an electric field applied along the $x$ direction and $|J_x^{(2)}|$ induced by an electric field applied along the $y$ direction, respectively, for $E_x=E_y=0.01$ V\,\AA$^{-1}$ and $\tau=1$ ps~\cite{engineering}. The vertical dotted line marks the intrinsic Fermi level.}
    \label{fig:nlhcurrent}
\end{figure}

The corresponding nonlinear Hall current was evaluated from Eq.~(\ref{eq:NLHC}) using an applied electric field of $E_x=E_y=0.01~\mathrm{V/\AA}$ and a constant relaxation time of $\tau=1.0~\mathrm{ps}$ within the constant relaxation-time approximation~\cite{terada2025,Tsirkin2021,engineering}. Since the signed Hall current follows the same energy dependence as the nonlinear Hall conductivity, Fig.~\ref{fig:nlhcurrent} shows the magnitude of the nonlinear Hall current, which directly reflects the strength of the experimentally measurable nonlinear response. Consistent with the behavior of the nonlinear Hall conductivity, the pristine monolayer exhibits an almost negligible current, whereas uniaxial strain produces a pronounced enhancement over a broad energy range. The largest response is obtained at 2\% tensile strain, where $|J_y^{(2)}|$ reaches approximately $2.5\times10^{4}~\mathrm{A\,m^{-2}}$ near the valence-band edge, while larger strains broaden the current distribution with a reduced peak magnitude. Moreover, $|J_y^{(2)}|$ consistently exceeds $|J_x^{(2)}|$, reflecting the larger $\chi_{yxx}$ response and the anisotropic Berry curvature dipole induced by uniaxial strain. These results demonstrate that moderate tensile strain provides the optimum condition for maximizing the nonlinear Hall current and highlights the strong tunability of nonlinear transport in monolayer Janus AsTeBr.
\subsection{Strain-Tunable Optical Properties}
\begin{figure}
    \centering
    \includegraphics[width=1.0\linewidth]{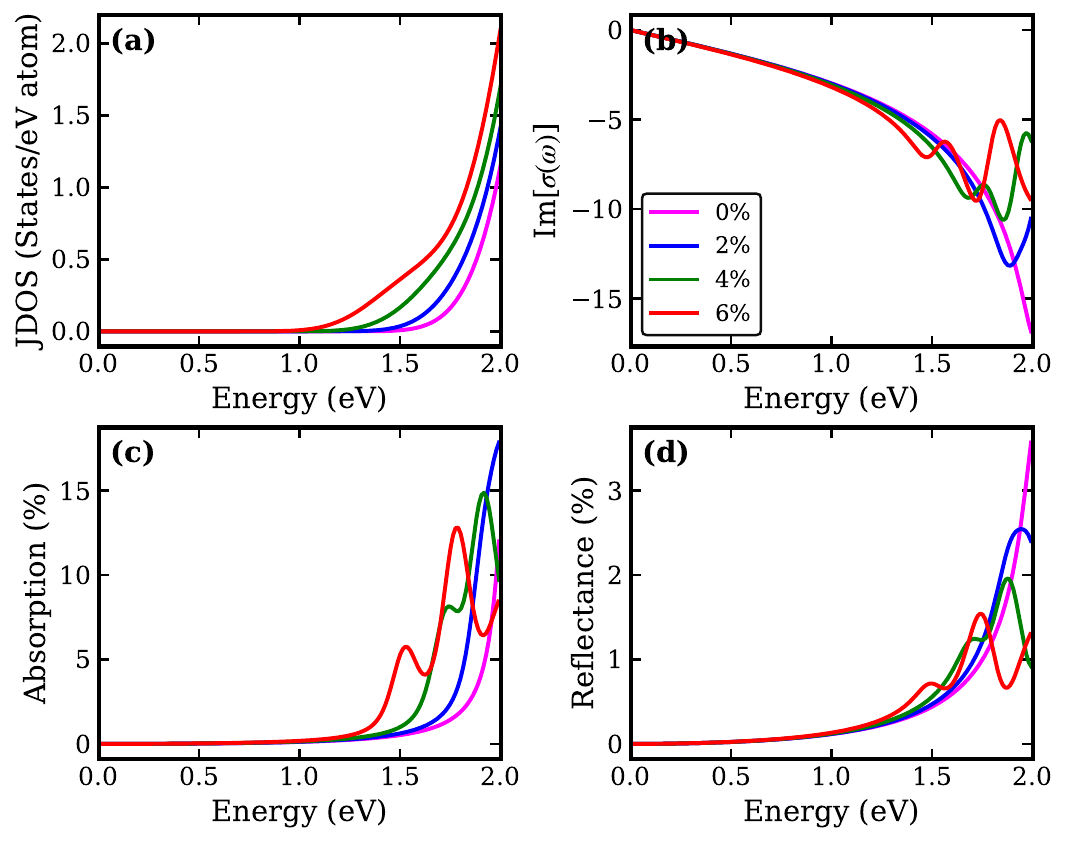}
    \caption{\justifying Strain-dependent optical properties of Janus AsTeBr under uniaxial strain (a) joint density of states (JDOS), (b) imaginary part of the dielectric function $Im[(\varepsilon(\omega))]$, (c) absorption spectrum, and (d) reflectance. The spectra show a progressive shift toward lower photon energies with increasing strain, consistent with the strain-induced modification of the electronic structure.
}
    \label{fig:optical}
\end{figure}
The strain-induced modification of the electronic structure is further reflected in the optical response of monolayer Janus AsTeBr. Fig.~\ref{fig:optical} presents the calculated joint density of states (JDOS), imaginary part of the dielectric function, optical absorption, and reflectance under different uniaxial tensile strains. Since these optical quantities originate from interband electronic transitions, they provide additional insight into the evolution of the strain-dependent band structure and orbital hybridization.

The calculated JDOS [Fig.~\ref{fig:optical}(a)] exhibits a clear shift of the absorption onset toward lower photon energies with increasing tensile strain, consistent with the reduction of the electronic band gap discussed in Sec.~III~A. The enhanced JDOS at lower energies indicates that strain increases the density of available interband transitions near the band edges. This behavior is also reflected in the imaginary part of the dielectric function, Im$[\varepsilon(\omega)]$, shown in Fig.~\ref{fig:optical}(b), where the principal optical features shift toward lower energies and evolve in intensity as the band dispersion and orbital hybridization are modified by strain. These changes originate from the strain-induced variation of the transition matrix elements between occupied and unoccupied electronic states.
The evolution of the dielectric response directly influences the optical absorption [Fig.~\ref{fig:optical}(c)]. The absorption edge undergoes a pronounced red shift with increasing tensile strain, accompanied by the emergence of stronger absorption peaks in the visible energy range. This enhancement arises from the increased probability of low-energy interband transitions resulting from the strain-induced narrowing of the band gap. A similar trend is observed in the reflectance spectra [Fig.~\ref{fig:optical}(d)], where the reflectance onset also shifts toward lower photon energies and the peak intensity evolves with strain, reflecting the modified dielectric response.

Although the optical response evolves monotonically with increasing tensile strain through the continuous reduction of the band gap, the nonlinear Hall response reaches its maximum at moderate strain owing to the nonmonotonic evolution of the Berry curvature dipole. This contrast highlights that the optical and nonlinear transport responses arise from related yet distinct consequences of strain-induced electronic structure reconstruction, providing multiple avenues for tuning the functionality of monolayer Janus AsTeBr.
\section{Conclusions}
In summary, we have systematically investigated the effect of uniaxial strain on the Berry-phase-driven transport and optical properties of monolayer Janus AsTeBr using first-principles calculations combined with Wannier-based transport theory. While pristine AsTeBr possesses finite local Berry curvature owing to broken inversion symmetry, its $C_{3v}$ crystal symmetry suppresses the BCD and consequently the intrinsic nonlinear Hall response. By lowering the symmetry to $C_{1}$ through uniaxial tensile strain, a finite BCD is generated via the anisotropic redistribution of Berry-curvature hotspots in momentum space.
Our calculations reveal that the BCD is strongly enhanced under uniaxial strain, reaching a maximum value of $0.20477$~\AA\ at 2\% tensile strain. This pronounced enhancement originates from the anisotropic redistribution of Berry-curvature hotspots induced by symmetry lowering and underpins the emergence of a robust intrinsic nonlinear Hall response. In parallel, uniaxial strain continuously tunes the optical response by red-shifting the absorption edge and modifying the dielectric function and reflectance through strain-dependent changes in the electronic structure.
These results establish the microscopic relationship between symmetry breaking, Berry-curvature redistribution, Berry curvature dipole formation, and nonlinear Hall transport in Janus AsTeBr. More generally, they demonstrate that symmetry engineering by uniaxial strain provides an effective strategy for controlling both Berry-phase-mediated transport and optical functionalities in noncentrosymmetric two-dimensional materials.
\section*{Acknowledgments}
The authors gratefully acknowledge the financial support provided by NISER Bhubaneswar. The calculations were performed on the KALINGA high-performance computing (HPC) cluster at NISER, Bhubaneswar.
\section*{DATA AVAILABILITY}
The data that support the findings of this article are not
publicly available. The data will be available from the authors
upon reasonable request.

\twocolumngrid
\bibliography{reference.bib}
\bibliographystyle{apsrev4-2}
\bibliographystyle{apsrev4-1}
\newpage
\section{supplementary Material}
\subsection{Structural Parameters}
The optimized structural and electronic parameters of monolayer Janus AsTeBr under uniaxial tensile strain applied along the $x$ direction are summarized in Table~\ref{tab:S1}. As the tensile strain increases, the lattice constant and the As--Te and As--Br bond lengths increase monotonically, while the electronic band gap decreases continuously owing to the strain-induced modification of orbital hybridization.
\begin{table}[H]
\caption{
Optimized structural and electronic parameters of monolayer Janus AsTeBr under uniaxial tensile strain applied along the $x$ direction. Bond lengths are given in \AA\ and energies in eV.
}

\centering
\renewcommand{\arraystretch}{1.2}
\setlength{\tabcolsep}{4pt}

\begin{tabular}{cccccccc}
\toprule
Strain &
$a_x$ &
$d_{\rm As-Te}$ &
$d_{\rm As-Br}$ &
$E_g$ &
VBM &
CBM &
$E_F$\\
(\%) & (\AA) & (\AA) & (\AA) & (eV) & (eV) & (eV) & (eV)\\
\midrule
0 & 3.86 & 2.80 & 2.95 & 1.527 & -2.268 & -0.741 & -2.190\\
2 & 3.94 & 2.84 & 2.98 & 1.448 & -2.399 & -0.951 & -2.282\\
4 & 4.02 & 2.88 & 3.01 & 1.325 & -2.519 & -1.193 & -2.440\\
6 & 4.09 & 2.92 & 3.05 & 1.185 & -2.629 & -1.444 & -2.530\\
\bottomrule
\end{tabular}

\label{tab:S1}

\end{table}
\subsection{Phonon Dispersion}

\begin{figure}[H]
\centering
\includegraphics[width=0.95\linewidth]{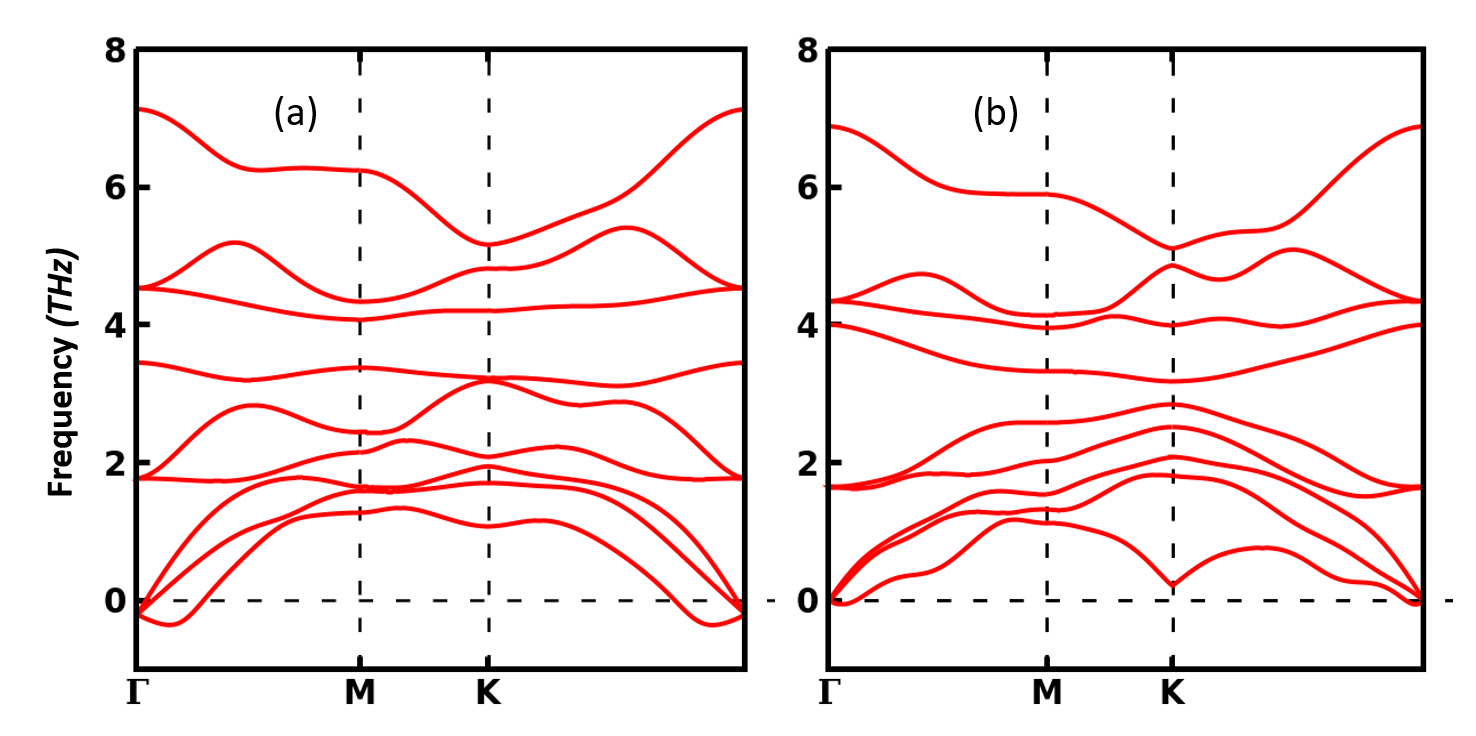}
\caption{
Calculated phonon dispersion of monolayer Janus AsTeBr under
(a) $-1\%$ compressive strain and
(b) $6\%$ tensile strain. Imaginary phonon modes appear under compressive strain, whereas the tensile-strained structure remains dynamically stable.
}
\label{fig:S1}
\end{figure}

To examine the effect of strain on the lattice stability of monolayer Janus AsTeBr, phonon dispersion calculations were performed within the framework of density functional perturbation theory (DFPT), as implemented in the PHONOPY package.\cite{phonopy1,phonopy2} Representative compressive ($-1\%$) and tensile (6\%) strain states were considered. As shown in Fig.~\ref{fig:S1}, imaginary phonon modes appear under $-1\%$ compressive strain, indicating dynamical instability. In contrast, the absence of imaginary phonon frequencies at 6\% tensile strain confirms the dynamical stability of the strained monolayer.


\subsection{Occupied Berry Curvature}
\begin{figure}[H]

\centering

\includegraphics[width=0.5\linewidth]{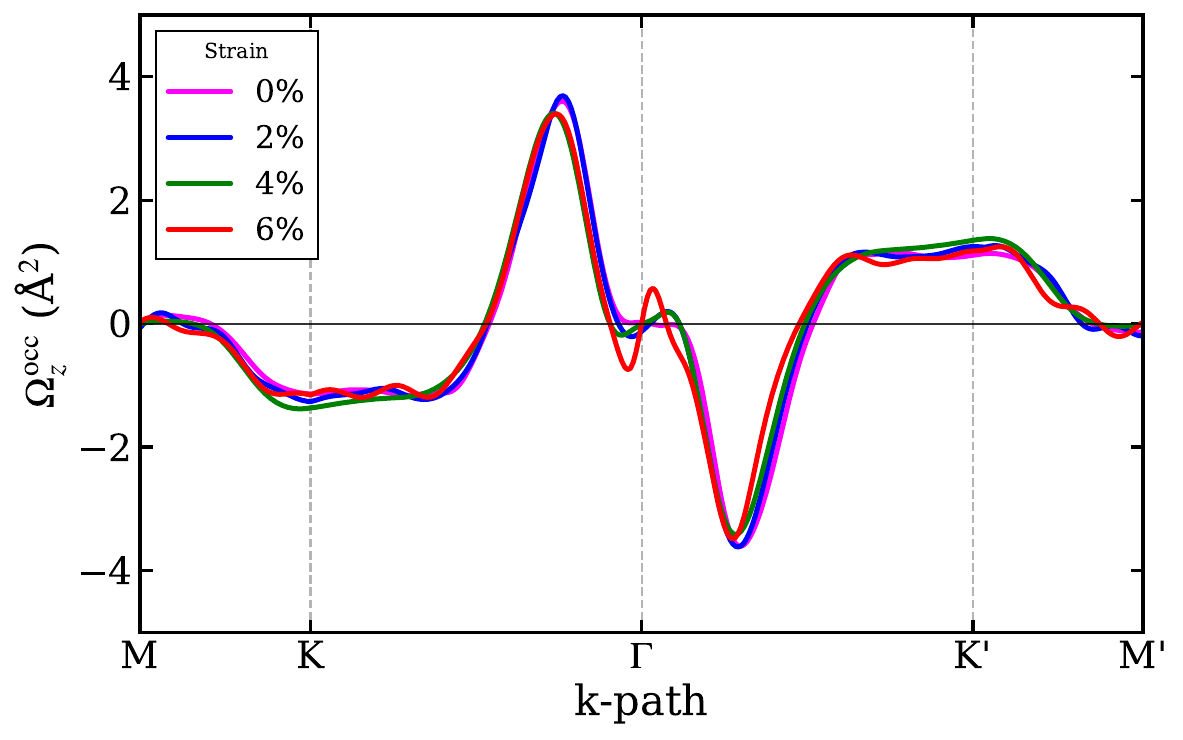}

\caption{
Occupied Berry curvature,
$\Omega_z^{\rm occ}$,
projected along the
$M$--$K$--$\Gamma$--$K'$--$M'$
high-symmetry path for
0\%, 2\%, 4\%, and 6\%
uniaxial tensile strain.
}

\label{fig:S2}

\end{figure}

Figure~\ref{fig:S2} presents the occupied Berry curvature projected along the high-symmetry path, obtained from Wannier interpolation.\cite{wannier90,wb1} Although the antisymmetric distribution required by time-reversal symmetry is preserved, uniaxial strain substantially modifies the magnitude and momentum-space distribution of the occupied Berry curvature~\cite{Sodemann2015}.
 The largest changes occur around the $\Gamma$ point, reflecting the strain-induced reconstruction of the electronic structure and providing the microscopic origin of the finite Berry curvature dipole.


\subsection{Berry Curvature Dipole}

 The Berry curvature dipole (BCD) was calculated using maximally localized Wannier functions constructed with Wannier90 and subsequently evaluated using the WannierBerri package.\cite{wannier90,wb1} The resulting tensor components are summarized in Table~\ref{tab:S2}. The pristine monolayer exhibits a vanishing BCD owing to the threefold rotational symmetry, whereas finite $D_{xz}$ and $D_{yz}$ components emerge upon application of uniaxial tensile strain  owing to symmetry lowering, giving rise to the nonlinear Hall response discussed in the main text. 


\begin{table}[H]

\caption{
Calculated Berry curvature dipole components $D_{xz}$ and $D_{yz}$ together with the corresponding magnitude $|D|$ for monolayer Janus AsTeBr under uniaxial tensile strain. All values are given in \AA.
}

\centering

\renewcommand{\arraystretch}{1.2}
\setlength{\tabcolsep}{12pt}

\begin{tabular}{cccc}

\toprule

Strain (\%) &
$D_{xz}$ (\AA) &
$D_{yz}$ (\AA) &
$|D|$ (\AA)\\

\midrule

0 & 0.00000 & 0.00 & 0.0000\\
2 & 0.13747 & 0.07 & 0.2047\\
4 & 0.08000 & 0.06 & 0.1400\\
6 & -0.07632 & 0.05 & 0.0263\\

\bottomrule

\end{tabular}

\label{tab:S2}

\end{table}
\subsection{Momentum-Space Distribution of the Berry Curvature Dipole}

To further illustrate the microscopic origin of the nonlinear Hall response, Fig.~S3 and Fig.~S4 present the momentum-space distributions of the Berry curvature dipole (BCD) integrand for representative tensile strains of 4\% and 6\%, respectively. The $D_x(\mathbf{k})$ and $D_y(\mathbf{k})$ components exhibit pronounced hot spots located primarily around the vicinity of the $K$ and $K'$ valleys. Owing to time-reversal symmetry, the distributions remain antisymmetric in momentum space, while the reduction of crystal symmetry under uniaxial strain leads to an unequal distribution of positive and negative contributions. The enhanced localization and increasing magnitude of these hot spots with increasing tensile strain reflect the strain-induced modification of the Berry-curvature distribution near the band extrema, which is responsible for the finite Berry curvature dipole and the resulting nonlinear Hall response.
\begin{figure}[H]
    \centering
    \includegraphics[width=0.95\linewidth]{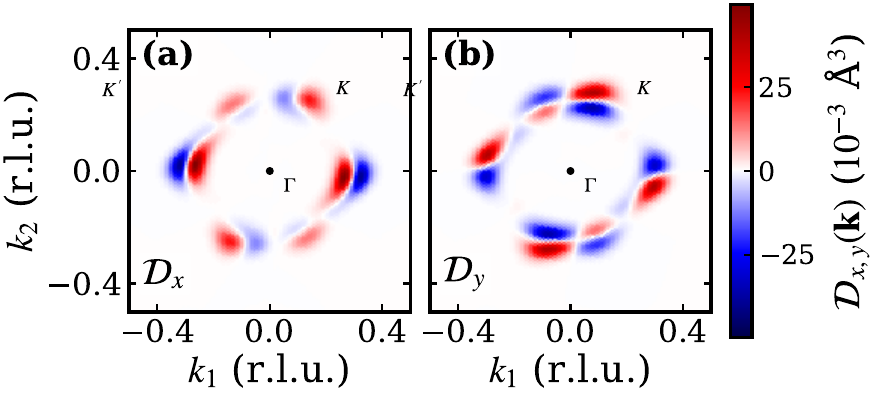}
    \caption{Momentum-space distribution of the Berry curvature dipole integrand for monolayer Janus AsTeBr under 4\% uniaxial tensile strain. Panels (a) and (b) show the $D_x(\mathbf{k})$ and $D_y(\mathbf{k})$ components, respectively. The largest contributions originate from the vicinity of the $K$ and $K'$ valleys.}
    \label{fig:S3}
\end{figure}

\begin{figure}[H]
    \centering
    \includegraphics[width=0.95\linewidth]{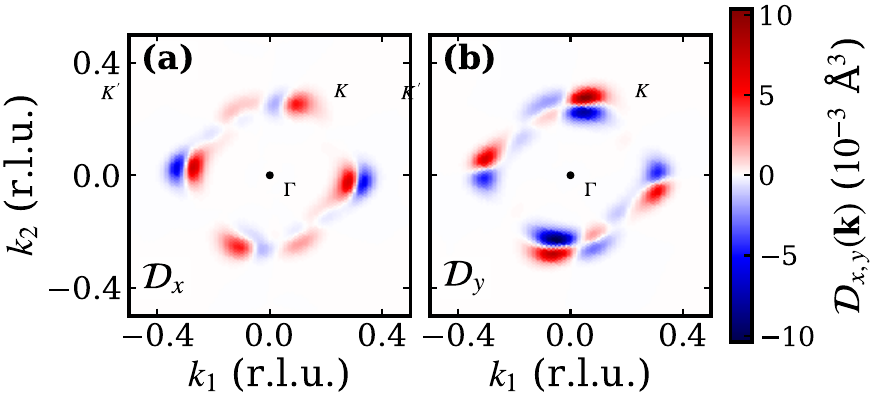}
    \caption{Momentum-space distribution of the Berry curvature dipole integrand for monolayer Janus AsTeBr under 6\% uniaxial tensile strain. The redistribution of positive and negative BCD hot spots under tensile strain reflects the evolution of the Berry-curvature landscape responsible for the nonlinear Hall response.}
    \label{fig:S4}
\end{figure}
\end{document}